\documentclass[a4paper,11pt]{article}
\usepackage{jinstpub} 
\usepackage{tikz}
\usepackage{caption}
\usepackage{subcaption}
\usepackage{cleveref}

\proceeding{AI4EIC 2023 Annual Workshop\\
Catholic University of America, Washington D.C.}

\title{Performance optimization for a scintillating glass electromagnetic calorimeter at the EIC}

\author[b]{J. Crafts}

\author[a]{R. Fatemi}

\author[b]{T. Horn}

\author[a,1]{D. Kalinkin,\note{Corresponding author.}}

\affiliation[a]{Department of Physics \& Astronomy, University of Kentucky,\\
  177~Chem.-Phys. Building, 506~Library Drive, Lexington, 40506-0055, KY, USA}

\affiliation[b]{Physics Department, American Christian University,\\
  620~Michigan Ave., N.E.~Washington, 20064, DC, USA}

\emailAdd{dkalinkin@uky.edu}

\abstract{
  The successful realization of the EIC scientific program requires the design
  and construction of high-performance particle detectors. Recent
  developments in the field of scientific computing and increased
  availability of high performance computing resources have made it
  possible to perform optimization of multi-parameter designs, even
  when the latter require longer computational times (for example
  simulations of particle interactions with matter). Procedures
  involving machine-assisted techniques used to inform the design
  decision have seen a considerable growth in popularity among the EIC
  detector community. Having already been realized for tracking and
  RICH PID detectors, it has a potential application in calorimetry
  designs. A SciGlass barrel calorimeter originally designed for EIC
  Detector-1 has a semi-projective geometry that allows for
  non-trivial performance gains, but also poses special challenges in
  the way of effective exploration of the design space while
  satisfying the available space and the cell dimension constraints
  together with the full detector acceptance requirement. This talk
  will cover specific approaches taken to perform this detector design
  optimization.
}


\definecolor{mlflowblue}{HTML}{1F77B4}
\definecolor{mlfloworange}{HTML}{FF7F0E}

\begin{document}
\maketitle
\flushbottom

\section{Introduction}
The Electron-Ion Collider (EIC), a new accelerator facility that will be built in the next decade, provides the novel opportunity to utilize Machine Learning algorithms directly in the accelerator, detector and data acquisition design. Previous efforts at optimizing particle detector designs for EIC included a dual radiator RICH detector \cite{jlab_drich} and a tracker subsystem \cite{ecce_tracker}, both were aimed for the ``Detector-I'' concept. Generally a design optimization problem is a Multi-Objective Optimization (MOO) problem with metrics which are practically non-differentiable. Such problems were successfully addressed in simpler cases using Genetic Algorithms, and there were recent developments in advanced methods for Bayesian Optimization (e.g., \cite{SAASBO}) showing promise for scaling to more complex tasks. The following paper summarizes findings in evaluation of the existing approaches for optimization of electromagnetic calorimeter detector to be used for EIC.

\begin{figure}[htbp]
\centering
\includegraphics[width=\textwidth,trim=0 0.1cm 0 0.1cm]{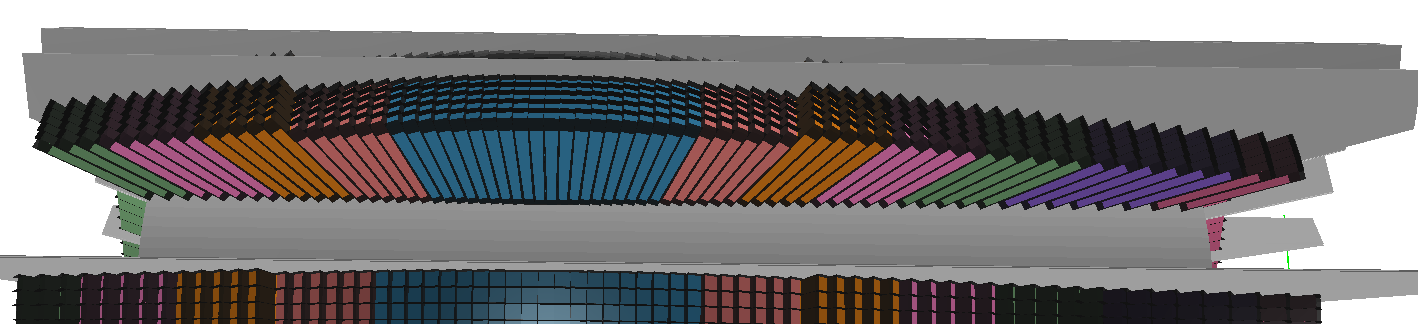}
\caption{Top-down view of one of the 12 sectors of the SciGlass projective geometry for Detector-II at EIC, with the sector in front of it removed for visibility. Seven different cell colors mark seven assumed shapes. Black denotes carbon fiber supports, and grey represent wall of the wedge box surrounding each sector.\label{fig:sciglass_design}}
\end{figure}

A projective homogeneous calorimeter with SciGlass radiator is envisioned for Detector-II at the EIC. This device will be used to measure the energy of electrons scattered at mid-rapidity (corresponding to interactions with high momentum transfer in deep-inelastic electron-proton scattering (DIS)). In this work, we attempt to solve the practical task of optimizing the geometrical shape of such a calorimeter. The reference design for this calorimeter with $\eta$-coverage $-1.7 < \eta < 1.3$  is shown on (\cref{fig:sciglass_design}).

\section{Problem definition}

\begin{wrapfigure}{r}{0.45\textwidth}
\centering
\vspace{-0.5cm}
\includegraphics[width=0.43\textwidth]{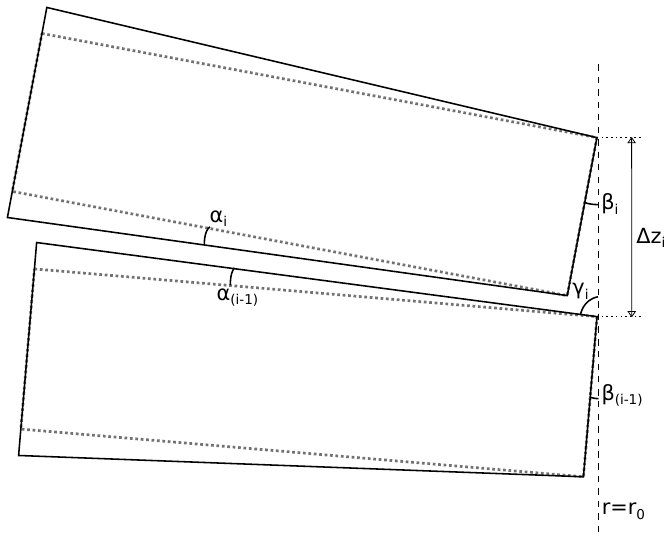}
\caption{Side-view schematic of longitudinal tower stacking with flaring angles $\{\alpha_i\}$.\label{fig:sciglass_tower_stacking}}
\end{wrapfigure}

The optimization procedure starts with defining a set of design parameters to be optimized and numerical objectives to quantify the detector performance. The focus of this work was on optimizing tower projectivity. To that end, the shapes of the towers were allowed to vary. The assumption was made that up to seven independent tower shapes can be manufactured, those are referred to as ``families''. The difference in shape was encoded in terms of flaring angles for each trapezoid that corresponded to the angle between opposite faces of a cell. When looking at the cells from the middle towards the detector ends that flaring angle would accumulate towards the polar angle of each cell's incline (\cref{fig:sciglass_tower_stacking}), since the adjacent sides of adjacent cells were coplanar with a 1~mm gap distance. An assumption was made that the towers from the same family would be stacked together, and families would go in the same order. That allowed to encode placements of the towers in an integer vector with 12 values: 7 for positive direction and 5 for the negative. Altogether, families flaring angles and numbers of towers used per family fit in a vector of 19 values.

The barrel SciGlass calorimeter subsystem serves several purposes: the measurement of electron and photon energy with excellent resolution, which is especially important for constraining kinematics of the scattered electron. It also needs to be able to help identify the scattered electron from the background of negatively charged hadrons, such as $\pi^-$ that will be also present in DIS events. This is largely relying on shower profiling and discrimination based on the ratio of the deposited energy ($E_{\text{dep.}}$), as reconstructed from a $3 \times 3$ tower cluster, and the momentum ($p_{\text{trk}}$) of the particle as determined by an external tracking detector.  In this work the true momentum of the particle ($p_{\text{thrown}}$) will be used since a realistic tracking detector has not been included in this study. Initial evaluations of pion rejection had showed that this design could use a slight improvement for particles at lower values of $p_T$ at which probabilities to encounter a pion are higher. Thus the focus of this work was on attempt at improving this quantity. Another responsibility of the subsystem is discrimination between $\gamma$ from DVCS and $\pi^0 \rightarrow \gamma\gamma$ signal from DVMP, which at high energies relies on computationally expensive ML training, and is less optimal for including in an initial round of optimization.

\section{Multi-objective Optimization using Genetic Algorithms}
\label{sec:NSGA}

A direct approach using NSGA-II \cite{nsga2} implemented in \texttt{pymoo} \cite{pymoo} framework was applied to the present problem. The algorithm displayed decent performance for the 2-objective problem when using default settings and population size of 100. One problem-specific consideration had to be made for the fact that implicit constraints placed by inherent possibility of overlaps occurring in the geometry, which prevent objective functions to be evaluated. The \cite{pymoo} framework does not allow for the user code to report such missing values, so instead values of 0 were reported when overlaps were detected, and non-overlapping minimized values were adjusted be always negative.

\section{Constraints and dimensional reduction}

The handling of constraints is particularly challenging for this problem. The cells of the calorimeter have to fit within the allocated volume, yet have a maximal possible acceptance within that limit. The resulting constraint on the parameters is non-linear due to complicated trigonometric relations arising for the angular parameters.

Another observation is that the time needed to evaluate the geometry for overlaps~(via~TGeo) is $\mathcal{O}(1\text{ second})$, much smaller than the time needed to evaluate the geometry for its performance which is at least $\mathcal{O}(1\text{ minute})$. One could ask then if it would be possible to pre-compute the manifold of parameters that correspond to valid geometries. The approach taken in this work is to use Markov Chain Monte Carlo (MCMC) walkers to explore the design space and approximately identify a subspace of valid geometries that occupy maximal acceptance within the detector envelope. The latter requirement is needed as we are not interested in valid geometries that don't use sufficient amount of towers. A value of the distance between the $z$-coordinates between the backward-most and forward-most towers is used as a proxy for acceptance that is easy to compute for a given geometry. In the end, the following probability distribution was given to the MCMC:

\begin{equation}
  \text{log}(P) = \left\{
  \begin{array}{l}
    -\infty\text{, if parameter set doesn't pass the overlap check}\\
    (z_{\text{rightmost tower}} - z_{\text{leftmost tower}}) / (1\text{ cm})\text{, otherwise}
  \end{array}
  \right.
\end{equation}

\begin{figure}[htbp]
  \centering
  \begin{subfigure}[b]{0.49\textwidth}
    \includegraphics[width=\textwidth]{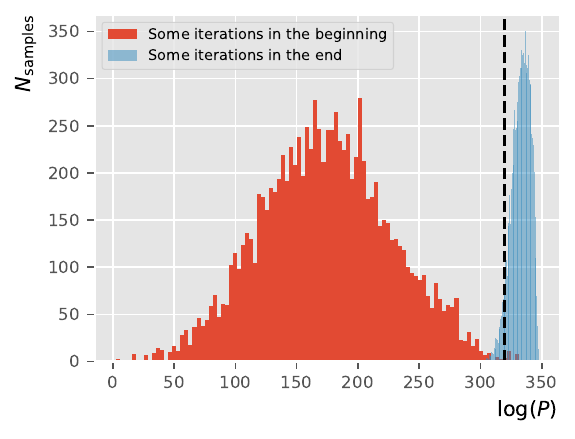}
    \caption{\label{fig:mcmc_results_logp}}
  \end{subfigure}
  \begin{subfigure}[b]{0.49\textwidth}
    \includegraphics[width=\textwidth]{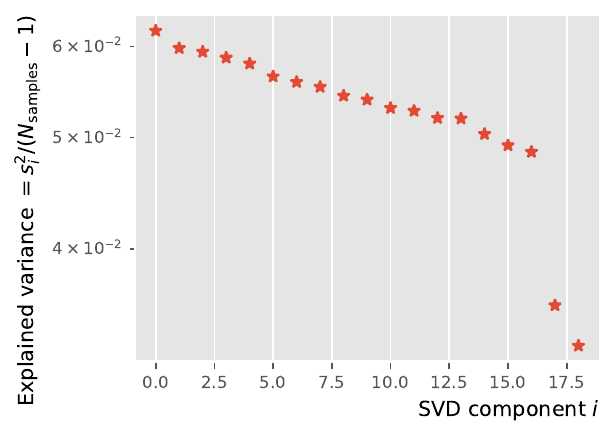}
    \caption{\label{fig:mcmc_results_svd_eig}}
  \end{subfigure}
  \\
  \begin{subfigure}[b]{0.49\textwidth}
    \includegraphics[width=\textwidth]{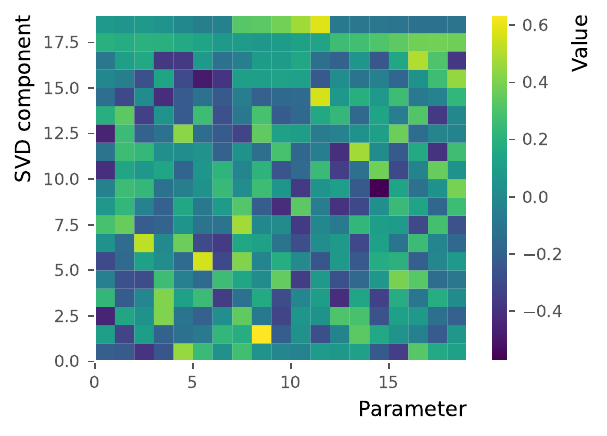}
    \caption{\label{fig:mcmc_results_svd_vec}}
  \end{subfigure}
  \begin{subfigure}[b]{0.49\textwidth}
    \includegraphics[width=\textwidth]{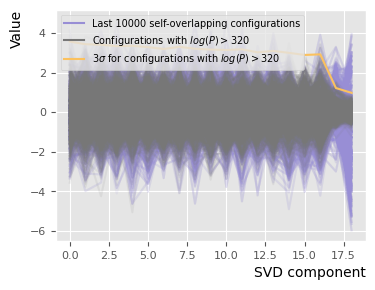}
    \caption{\label{fig:mcmc_results_corr}}
  \end{subfigure}
  \caption{Results from MCMC, including analysis using PCA.\label{fig:mcmc_results}}
\end{figure}

The MCMC was run until it converged to a concrete population effectively sampling a region of high $\text{log}(P)$, as seen on \cref{fig:mcmc_results_logp}. This population was then used to perform dimensional reduction. Generally, that would be achieved using a manifold learning method, but in this case, a simple Principal Component Analysis was applied. The main outcome was that two dimensions were highly constrained, as seen on \cref{fig:mcmc_results_svd_eig}. The eigenvectors corresponding to those directions, as seen on \cref{fig:mcmc_results_svd_vec}, had comparable components for variations in parameters of the same type. The PCA defines a transformation to a new parameter set. The limits for transformed parameters are not well-defined anymore, so instead a $3\sigma$ variation across populations was used. \Cref{fig:mcmc_results_corr} shows how variations are different between initial population and converged population after MCMC. It makes sense that there were two components reduced, as when detector geometry is defined by stacking from the center in the $\pm \hat{z}$ directions, there are two detector envelope boundaries to hit.

\begin{figure}[htbp]
  \centering
  \begin{tikzpicture}
    \node (plot) {
      \includegraphics[width=0.6\textwidth,clip,trim=200pt 500pt 100pt 150pt]{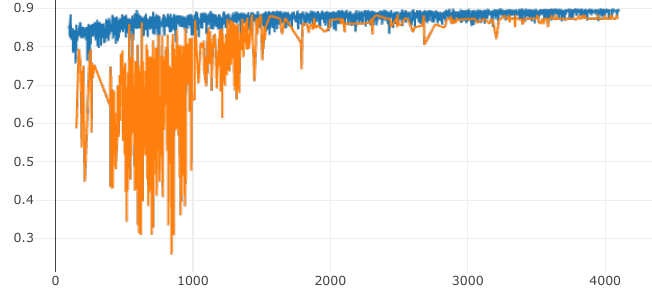}
    };
    \node (xlabel) at (1.,-2.3) {\small Generation * 100 + Sample index};
    \node[rotate=90] (ylabel) at (-4.7,1.2) {\small $E_{\text{dep.}} / p_{\text{thrown}}$};
    \node[align=right] (legend) at (1,0) {\color{mlfloworange}Without manifold\\\color{mlflowblue}With manifold};
  \end{tikzpicture}
  \caption{Values of energy deposition fraction at a given sample and generation for two-objective optimization using NSGA-II for raw parameterization vs parameterization in transformed coordinates.\label{fig:dim_reg_result}}
\end{figure}
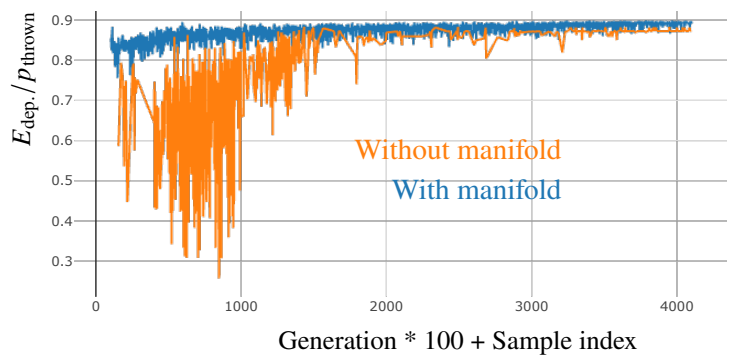

The effect of dimensional reduction was evaluated for 2-objective MOO using NSGA-II implementation described in \cref{sec:NSGA}. In comparison to optimization without dimensional reduction the convergence was achieved much faster with a slightly better outcome for the $E_{\text{dep.}} / p_{\text{thrown}}$ objective as illustrated in \cref{fig:dim_reg_result}.

\section{Multi-objective Optimization using Bayesian Optimization}
\label{sec:BO}

Bayesian Optimization (BO) is another popular approach for doing MOO. It has the benefit of reducing a number of expensive objective evaluations compared to Genetic Algorithms by reducing exploration and increasing exploitation. This approach however generally requires extra fine tuning to make it work.

The implementation of BO in the \texttt{Ax} framework with SAASBO \cite{SAASBO} surrogate model in the qNEHVI \cite{qNEHVI} acquisition function was used in this study. The surrogate model was initialized by fitting it to evaluated designs with parameters from a Sobol quasi-random sequence of length 1000. That was followed by several hundred of BO iterations for which $q=3$ samples were evaluated at a time and the surrogate model was refitted at each iteration. Unlike for NSGA-II the computational overhead of the optimization algorithms was not negligible compared to the cost of evaluating the objective functions. Instead, it was a dominant cost in the time budget of the whole procedure.

The same overlap consideration from \cref{sec:NSGA} applies to implementation in \texttt{Ax}. However, in this case the missing values must also be ignored  when determining a surrogate model fit model. The solution of returning a value of 0 with a large uncertainty appears to work sufficiently well for this purpose.

\begin{figure}[htbp]
  \centering
  \begin{subfigure}[b]{0.49\textwidth}
    \includegraphics[scale=0.9]{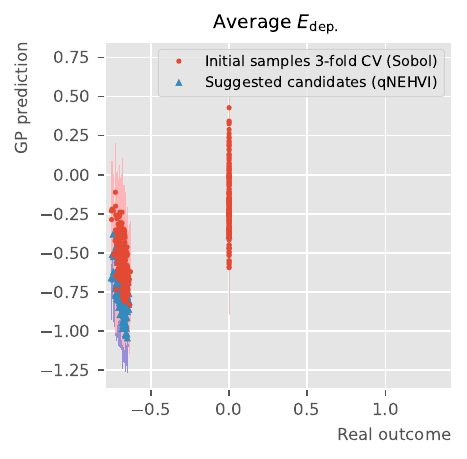}
    \caption{}
  \end{subfigure}
  \begin{subfigure}[b]{0.49\textwidth}
    \includegraphics[scale=0.9]{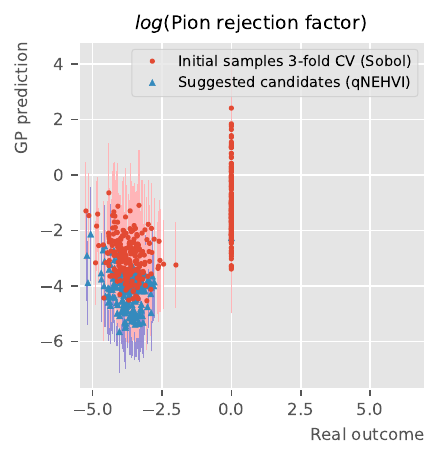}
    \caption{}
  \end{subfigure}
  \caption{Cross-validation for Gaussian Process surrogate models. Red dots mark 3-fold cross validation of the Sobol-generated sample within the fit and blue trianlges mark predicted and actual values for the points suggested by qNEHVI.\label{fig:cv}}
\end{figure}

Another important consideration for BO lies within the surrogate model's ability to adequately model the unknown objective function. To ensure that, a 3-fold cross validation has been performed for each of the objectives (shown on \cref{fig:cv}) on points from a Sobol-generated sample. The cross validation shows decent correlation, especially for pion rejection factor\footnote{In fact, the performance in cross validation can be improved slightly by tweaking the GP kernel. Unfortunately that one is hard-coded in the default implementation of the ``\texttt{FULLYBAYESIANMOO}'' model in the version of \texttt{Ax} available at the time of this writing.}. Furthermore, correlation of predicted versus actual objective value can be visualized by the points suggested by qNEHVI. Those points, by construction, lie lower than typical predictions for the Sobol model, but they also often give lower actual outcomes, which indicates that qNEHVI with the given surrogate model is capable at picking improving points.

An attempt has been made at utilizing the \texttt{OutcomeConstraint} facility in \texttt{Ax} in hope of reducing chances of picking the invalid combinations of the parameters by requiring number of overlaps to be $\leq 0$. The way such constraint is handled in \texttt{Ax} is that it receives a surrogate model of its own, however, the cross validation for that was not satisfactory. In practice, running optimization without this constraint, like for GA, worked sufficiently well.

\section{Software stack}

As was explained earlier, \texttt{pymoo} and \texttt{Ax} frameworks were used for optimization. The detector geometry description was implemented for SciGlass calorimeter within \texttt{DD4hep} framework~\cite{dd4hep} in which the evaluation of different designs was achieved by automatically producing alternative ``compact'' XML file configurations with updated numerical parameters. The job scheduling was performed using \texttt{Dask.Distributed} cluster with workers running using a \texttt{Slurm} batch system. The development of the software was facilitated by caching evaluations using \texttt{joblib.Memory} memoization that performed well when accessing a common cache situated on network storage despite concurrent access from multiple nodes. The results of individual experiments were tracked using \texttt{MLflow}. The simulation output produced by \texttt{DD4hep}'s interface to the \texttt{Geant4} (\texttt{ddsim}) was analyzed using implementation of objective functions using \texttt{Awkward Array} \cite{awkward} in-memory data representation loaded using \texttt{uproot5} \cite{uproot5}.

\section{Results}

\begin{figure}[htbp]
\centering
  \begin{subfigure}[b]{0.49\textwidth}
    \includegraphics[width=\textwidth]{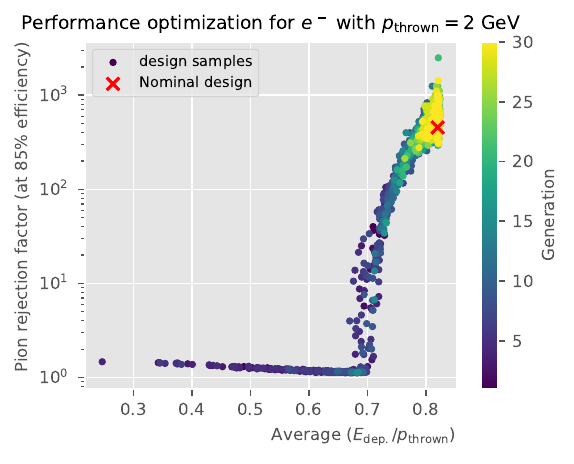}
    \caption{\label{fig:result_ga}}
  \end{subfigure}
  \begin{subfigure}[b]{0.49\textwidth}
    \includegraphics[width=\textwidth]{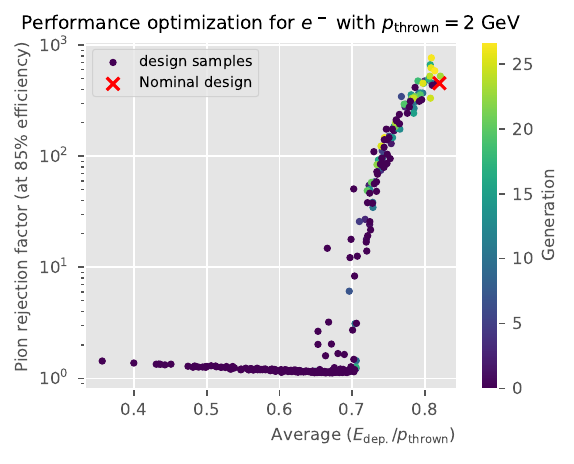}
    \caption{\label{fig:result_bo}}
  \end{subfigure}
  \caption{Objective points from 40 iterations of NSGA-II (a) and Sobol and Bayesian Optimization (b). The red cross indicates the reference objective point corresponding to a detector configuration using hand-picked flaring angles and per-family tower counts.}
\end{figure}

\begin{figure}[htbp]
\centering
\includegraphics[width=0.50\textwidth]{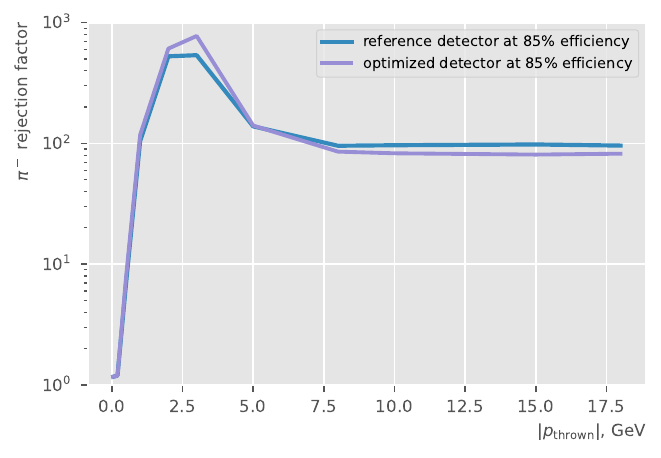}
\caption{Dependencies of pion rejection factors on thrown charged particle transverse momentum for the best NSGA-II result.\label{fig:benchmark_ga}}
\end{figure}

Optimization using both Genetic Algorithms (GA) and Bayesian Optimization (BO) approaches were performed for the SciGlass detector targeting two objectives: $E_{\text{dep.}} / p_{\text{thrown}}$ (proxy for the detector acceptance and, partially, for the energy resolution) for $2$~GeV electrons.

The result for GA is shown on \cref{fig:result_ga}. The narrow shape of the correlation suggests, in hindsight, that the optimization was close to being de-facto single-objective. The optimized detector was able to outperform the reference one in terms of pion rejection factor by a half of an order in magnitude. Full evaluation using a benchmarking software from Detector-I technology review \cite{SciGlass_benchmark} showed (\cref{fig:benchmark_ga}) how the optimal configuration trades diminished pion rejection at high transverse momentum for increase at low transverse momentum. This is a desirable trade as pion contamination is lesser for electron candidates with a high momentum.

\begin{figure}[htbp]
\centering

  \begin{tikzpicture}
    \node (plot) {
      \includegraphics[width=0.6\textwidth,clip,trim=200pt 500pt 100pt 140pt]{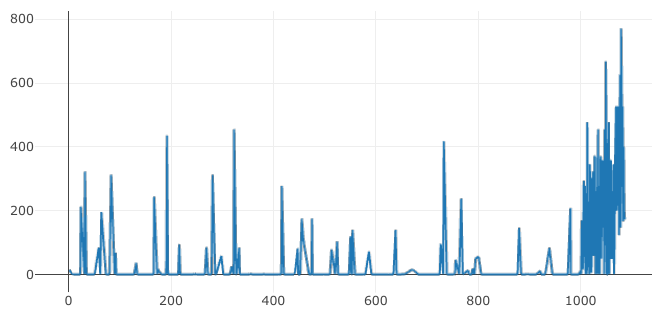}
    };
    \node (xlabel) at (3.4,-2.4) {\small Sample index};
    \node[rotate=90] (ylabel) at (-4.8,0.8) {\small $\pi^-$ rejection factor};
  \end{tikzpicture}
\caption{Pion rejection as a function of iteration. Where first 1000 entries are the Sobol quasi-random sequence and what follows are the actual Bayesian Optimization iterations.\label{fig:performance_bo}}
\end{figure}

The result for BO shown on \cref{fig:result_bo} closely resembles the one obtained from the GA method, however shows that a potentially larger runtime, and, possibly, also additional setup, would be required to achieve benchmark performance of the GA method. \Cref{fig:performance_bo} shows real-time performance of the BO method, which picks up some local optimum after about $\approx 100$ evaluations (corresponding to the iterations $\approx 1100$ and further on the plot).

\section{Conclusion}

This work demonstrates an example of optimization of a real world projective calorimeter design for the future EIC Facility. The SciGlass detector has a potential application as a prominent design mid-rapidity electron measurement device in future Detector-II, and, as was demonstrated, can be further bettered using ML optimization techniques. The described suggested method for handling of implicit parameter constraints should be applicable for a wider range of problems, including to the problem of integrating dimensions of subsystems in full detector optimization tasks.

\acknowledgments

We would thank the University of Kentucky Center for Computational Sciences and Information Technology Services Research Computing for their support and use of the Lipscomb Compute Cluster and associated research computing resources.



\bibliographystyle{JHEP}


\end{document}